\documentclass[]{interact}

\usepackage[square,numbers]{natbib}
\usepackage{amsmath,amssymb,amsfonts}
\usepackage{graphicx}
\usepackage{textcomp}
\usepackage{xcolor}

\usepackage{algcompatible}
\usepackage{algorithm}

\usepackage{soul}

\newcommand{\bfepsilon}{{\hbox{\boldmath$\epsilon$}}}
\newcommand{\bfx}{{\mathbf{x}}}
\newcommand{\bfT}{{\mathbf{T}}}
\newcommand{\bfS}{{\mathbf{S}}}

\newcommand{\bft}{{\mathbf{t}}}

\def\BibTeX{{\rm B\kern-.05em{\sc i\kern-.025em b}\kern-.08em
    T\kern-.1667em\lower.7ex\hbox{E}\kern-.125emX}}

\begin{document}
%
\title{Real-Time Anomaly Detection for Advanced Manufacturing: Improving on Twitter's State of the Art}%
%
%
%

\author{Caitr\'iona M. Ryan,
       Andrew C. Parnell,
        and Catherine Mahoney
\thanks{\textit{E-mail: triona.ryan@mu.ie}}
\thanks{C.M. Ryan and A.C. Parnell are with I-Form / Hamilton Institute, Maynooth University, Ireland} 
\thanks{C. Mahoney is with University College Dublin, Ireland.}
\thanks{Manuscript received July 21. 2020.}
}

\maketitle 

\begin{abstract}
The detection of anomalies in real time is paramount to maintain performance and efficiency across a wide range of applications  including web services and  smart manufacturing. This paper presents a novel algorithm to detect anomalies in streaming time series data via statistical learning. Using  time series decomposition, a sliding window approach and recursive updates of the test statistics,   the generalised extreme studentised deviate (ESD) test is made feasible for streaming time series data.  
 The method is statistically principled and it outperforms the \texttt{AnomalyDetection}  software package, recently released by Twitter  Inc. (Twitter)  and used by multiple teams  at Twitter as their state of the art on a daily basis. The methodology is demonstrated using unlabelled data from the Twitter \texttt{AnomalyDetection} GitHub repository,  a manufacturing example with labelled anomalies and the Yahoo EGADS benchmark data.\end{abstract}


%
\maketitle

\section{Introduction}\label{sec:intro}
The detection of anomalies  (data that deviates from what is expected) is important to protect revenue, reputation and resources in many applications such as web services, smart manufacturing, telecommunications, fraud detection and biosurveillance. For example, exogenic factors  such as bots, spams and sporting events can affect web services  as can hardware problems and other endogenic factors \cite{twitter17}. 
In advanced manufacturing, the detection of anomalies in streaming machine data, for example from machine sensors that monitor processing conditions, can aid the identification of tool wear and tear and any problems in  the structure or quality of a part in production  \cite{konrad18}. 
Figure $1$ displays one such time series with labelled anomalies taken from the Numenta data repository \cite{lavin15}; temperature sensor data from an internal component of a large, industrial machine. 
 The first anomaly is not explained or discussed in \cite{lavin15}. The second anomaly is a planned shutdown of the machine. The third anomaly is difficult to detect and directly led to the fourth anomaly, a catastrophic failure of the machine.  
\begin{figure}[!t]
\centering
\includegraphics[width=2.5in]{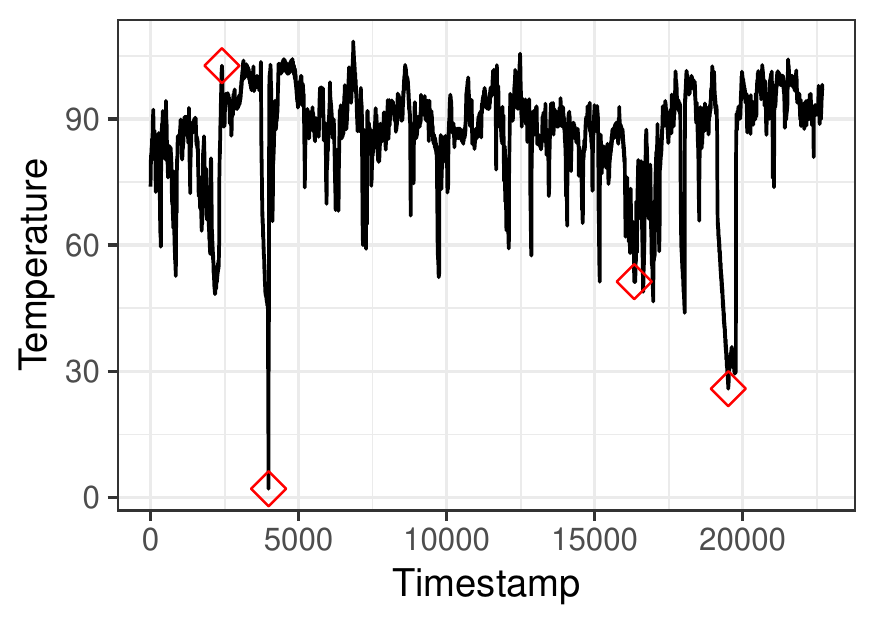}
\caption{Machine Temperature data stream with known anomalies marked in red} 
\label{fig:intromachines}
\end{figure}

The \texttt{AnomalyDetection} software package \cite{twitterpackage} was recently released by Twitter and is used daily to detect anomalies in their cloud infrastructure data, for example Tweets Per Second (TPS) and CPU utilisation.  A conference publication \cite{twitter14}, an article  published on ArXiv \cite{twitter17} and a blogpost \cite{twitterblog15} have generated much interest with over $120$ citations since 2014, accepting Twitter's challenge to the public and academic community to ``evolve the package and learn from it as they have'' \cite{twitterblog15}. 

The problem of anomaly detection in time series data of this nature is challenging due to its seasonal nature and its tendency to exhibit a trend. 
The approach taken by Twitter \cite{twitter14,twitter17,twitterblog15,twitterpackage} is the Seasonal Hybrid Extreme Studentised Deviate  (SH-ESD) algorithm. This is an adaptation of the generalised extreme studentised deviate (ESD) test \cite{rosner1983} which is itself a repeated application of  the Grubbs hypothesis test \cite{grubbs1950}  for a single outlier. These tests assume that the data is normally distributed. Thus it is necessary  to decompose the time series, subtract the seasonal and trend components and perform the hypothesis tests on the resulting residuals. 
In order to perform the decomposition, SH-ESD \cite{twitter14,twitter17,twitterblog15} uses the median value of non-overlapping windows of data  to estimate the trend as a stepwise function,  which they argue is more robust to outliers.
Whilst this is computationally fast and works well for some datasets, the results can be especially sensitive to the choice of window size and location.
SH-ESD uses LOESS  to  determine the seasonal component via  \texttt{stl} \cite{stlpackage}. However, SH-ESD requires the period to be specified by the user and a further requirement is that each non-overlapping window is assumed to capture at least one period of each seasonal pattern. In contrast, the  approach presented in this paper avoids  these restrictions and is applied to a rolling window of streaming data. 

A major statistical consideration arises from the implementation of
SH-ESD.
In order to increase the robustness of the method against a large number of outliers, SH-ESD uses the median and median absolute deviations (MAD) to studentise observations. This is not statistically appropriate for the generalised ESD test. 
The ESD test is derived under a Central Limit Theorem assumption of normalisation by the mean and standard deviation. The resulting critical value approximation due to \cite{rosner1975} accounts for the presence of outliers and thus does not require robust statistics such as median and MAD. The use of median and MAD for studentising observations may result in values that follow a heavier-tailed distribution than the adjusted t-distribution used in the significance tests.  This is a potential cause of high levels of type I error when using SH-ESD. 
 
This paper presents a novel  algorithm entitled Recursive ESD  (R-ESD) for fast anomaly detection in time series streaming data. It takes a statistically principled approach and addresses  the problems with SH-ESD outlined above.  Formulating the test statistic in a novel recursive way allows the test to be implemented while streaming data in real time. This is a key improvement over existing methods, enabling a real-time, low memory, statistically principled version of a widely used approach.
First,  the seasonality and trend is estimated in an initial phase. Then the statistically principled generalised ESD test is implemented in a sliding window of time series data.   
The R-ESD approach results in fast identification of anomalies in each window which can be communicated to the end user while the data is being streamed. We address the problems in SH-ESD outlined above by; (i)  formulating two recursive updates of the ESD test statistic,
enabling anomaly detection while streaming in real time;  (ii) using the mean to studentise observations, as it is statistically principled and appropriate for the statistical hypothesis test for outliers and (iii) estimating the period in the initial phase of the algorithm using a Fourier transformation via the \texttt{periodogram} function in TSA \cite{tsapackage}.

R-ESD is statistically principled. Moreover, SH-ESD has a number of further limitations where our method adds real value;
(i) SH-ESD use non-overlapping windows of data. Thus its nature is prohibitive for true streaming data. R-ESD detects anomalies while streaming;
(ii)  In SH-ESD, each non-overlapping window is assumed to capture at least one period of each seasonal pattern. R-ESD is fast and easy to apply to a new dataset with no prior knowledge;
(iii) In SH-ESD, this period must be pre-specified. This is not so for R-ESD;
 (iv) SH-ESD is prone to high levels of type one error (the detection of false positive anomalies), perhaps due to the incorrect use of median and MAD as described above. R-ESD outperforms SH-ESD and is statistically principled.

Other approaches to the problem of anomaly detection in times series data include Yahoo Inc.'s anomaly detection system; EGADS: the Extensible Generic
Anomaly Detection System \cite{laptev2015egads},  DeepAnT \cite{munir2018DeepAnT} and DeepAD \cite{buda2018deepad}. The latter two are acronyms of Deep Learning based Anomaly Detection methods. 
 The strength of EGADS is in combining many forecasting models and anomaly detection methods to detect as many different types of anomalies as possible. As \cite{laptev2015egads} admits, there is no best anomaly detection model for all use cases and certain algorithms are best at detecting different types of anomalies. Using EGADS, SH-ESD performed best on A3 (also known as TS-3) (see Figure $5$ of \cite{laptev2015egads}), which includes outliers but no change points. Section \ref{sec:results}  demonstrates that R-ESD outperforms SH-ESD, thus it also outperforms EGADS for this dataset.
DeepAnT (Munir et al., 2018) and DeepAD (Buda et al., 2018) are threshold based whereas R-ESD uses a principled statistical hypothesis test to discern an anomaly.  
DeepAnT uses convolutional neural network models (CNN) to predict the next value of a rolling window of time series data. A threshold on the Euclidean distance from actual to predicted value is used  for the anomaly detection. 
DeepAD combines the predictions of multiple forecasting techniques to forecast each newly streamed datapoint and detect anomalies using a dynamic extreme value threshold.
In both DeepAnT and DeepAD, one or multiple models are fit at every newly streamed datapoint whereas in R-ESD one model is fit for the entire dataset. In this sense the most comparable method to R-ESD is DeepADVote, a subset of DeepAD that fits only one model. This article demonstrates that R-ESD outperforms DeepADVote in F1-Score for the Yahoo EGADS Webscope benchmark A3 data  \cite{webscopedata}. Further comparisons and discussion of the benefits of R-ESD over and above its competitors will be demonstrated in Section \ref{sec:results}.

The paper is organised as follows.  Section \ref{sec:AD} describes the problem of anomaly detection, including a short review of existing methods and outlines how we propose to measure the performance of R-ESD. Section  \ref{sec:grubbs} formally defines the problem and presents the R-ESD method. In Section \ref{sec:results}, the R-ESD approach is demonstrated using the manufacturing dataset displayed in Figure $1$, using unlabelled data from the Twitter \texttt{AnomalyDetection} GitHub repository \cite{twitterpackage}  and using the Yahoo EGADS Webscope  A3 benchmark data \cite{webscopedata}.
We conclude with a discussion in Section \ref{sec:discussion}.

\section{Anomaly detection}\label{sec:AD}

Anomaly detection is the problem of identifying patterns in data that
do not conform to typical behaviour. By definition, the anomaly detection problem depends on the data and or application in question. 
 \cite{chandolaetal09} provide a thorough review and compare a range of approaches to anomaly detection given in the scientific literature including examples from industrial damage detection, medical anomaly detection,  cyber intrusion detection and sensor networks. \cite{gupta14}  provide a survey of anomaly detection methods in the computer science literature for temporal data. 
The challenge of selecting a suitable algorithm is discussed in  \cite{kandanaarachchi18}. \cite{leighetal19} presents a framework to identify and compare suitable methods for a water-quality problem to detect anomalies in high frequency sensor data.

In this paper we focus on the problem of univariate time series streaming data such as those  presented by Twitter \cite{twitter14,twitter17,twitterblog15} and the manufacturing problem displayed in Figure \ref{fig:intromachines}. This problem is of major importance, evidenced by the \texttt{anomalydetection} package, \texttt{tsoutliers}, \texttt{anomalize} and other highly cited packages being similarly focused.
The typical behaviour of data of this nature is to exhibit trend and/or seasonality. The existence of a trend might itself be an anomaly. The research challenge is to detect anomalous data points as they arrive in streaming applications or soon after they arrive.    The exact nature of the streaming capacity will be application dependent.

Consider a univariate time series data stream $\ldots,x_{t-1}, x_t, x_{t+1} \ldots$, where $x_t$ is an observation recorded at time $t$. We implement a rolling window approach to the problem of anomaly detection, where the window $\bfx = \{x_{t-w}, x_{t-w+1}, \ldots , x_{t} \}$ is the set of the $w$  observations of the data stream  prior to and including time $t$. 
The goal is to detect anomalies $\tilde{\bfx} \in \bfx$ and $\tilde{\bft} \in \{t-w,\ldots, t\}$, the anomalies and associated time-stamps of a subset of observations in each rolling window while streaming. 

In order to assess an anomaly detection algorithm, it is useful to analyse datasets that are annotated with ground truth labels.  In such cases, one can measure the precision; the proportion of true positive anomalies of all detected anomalies and recall; the ratio of true positive anomalies to the sum of true positive anomalies and false negative anomalies.  We use these measures to compare our performance to the Twitter \texttt{AnomalyDetection} package in detecting known anomalies in manufacturing data.

\section{ Recursive ESD for  streaming time series data}\label{sec:grubbs}

 The Grubb's test provides a hypothesis test for a single outlier \cite{grubbs1950}.
This was generalised to the ESD test \cite{rosner1983}, where a pre-specified number of $k$ anomalies can be detected. The ESD test statistics $R_1,\ldots,R_k$ are calculated from  samples of 
size $n,n-1,\ldots,n-k+1$, successively reduced  by the most extreme deviate (and potential anomaly) in the sample. 
For example, in the full sample of size $n$, the most extreme deviate would correspond to $x_i$, such that $\| x_i - \bar{x} \| \ge \| x_j- \bar{x} \| \; \;\; \forall i,j = 1, \ldots, n$, with equality only when $i=j$. $\bar{x}$ is the full sample mean.
This is computed analogously for subsequent reduced samples.

In general, we denote $\tilde{x}_j$ as the dataset with the $j$th most extreme deviates removed and $\tilde{n}_j$ as the sample size of this set. So, for example $\tilde{n}_1$ will be equal $n - 1$ when the most extreme deviate is removed. The ESD test statistic is defined by
\begin{equation*}
R_{j+1}=\frac{\max_i \left|x_i-\bar{\tilde x}_{j}\right|}{\tilde s_j}, \;\;\; i=1,\ldots, \tilde n _ {j};  \;\;\; j=1,\ldots, k;
\end{equation*}
where the reduced sample mean is
\begin{equation*}
\bar{\tilde x}_{j} = \frac{\sum_{i=1}^{\tilde n _ {j}}\tilde {x}_i}{\tilde n _ {j}},
\end{equation*} 
and where the sum of squares with all the $j$th most extreme deviates removed is

\begin{equation*}
\tilde s^2_j = \frac{\sum_{i=1}^{\tilde n _ {j}}\left(x_i-\bar {\tilde x}_{j}\right)^2}{\tilde n _ {j}-1}.
\end{equation*}

 The critical values for this series of Student's t-tests are
\begin{equation}
\gamma_{l+1} = \frac{t_{n-l-2,p} (n-l-1)}{\sqrt{(n-l-2+t^2_{n-l-2,p}) (n-l)}}, \hspace{0.2cm} l=0,1,\ldots, k-1,
\end{equation}
where $n$ is the number of data points in the dataset, $k$ is the maximum number of anomalies, $l$ is the order statistic and $p = 1 - (\alpha/2)(n-l)$.
Further details can be found in  Equation $2.5$ of  \cite{rosner1983}. 

In order to adopt the ESD test for streaming data, we note that the ESD test statistic $R_{j+1}$ can also be expressed as a function of the Grubb's ratio $\frac{S_n^2}{S^2}$ used in  \cite{grubbs1950} such that, in our notation;  
\begin{equation}
R_{j+1} = \sqrt{\left(1-\frac{\tilde S^{2}_{j+1} }{\tilde S^{2}_{j}}\right)(\tilde n _ {j}-1)},
\end{equation}
where
\begin{equation*}
\frac{\tilde S^2_{j+1}}{\tilde S^2_{j}}= \frac{\sum_{i=1}^{\tilde n _ {j+1}} \left(x_i-\bar {\tilde x} _{j+1} \right)^2}{\sum_{i=1}^{\tilde n _ {j}} \left(x_i-\bar{ \tilde x }_{j}  \right)^2}. 
\end{equation*}
This construction of the ESD test statistic is novel and useful in the context of our streaming anomaly detection problem as it permits recursive calculations. Having identified $x^* $ as the most extreme deviate in a sample $\tilde x$, the sums of squares can be reduced using the following recursive calculation;
\begin{equation}
\tilde {S}_{j+1}^2=\tilde {S}_{j}^2 - \tilde{ n}_{j+1} (x^* - \bar {\tilde{x}}_j )^2 /\tilde{n}_j.
\end{equation}
The recursive ESD test  is outlined in Algorithm 1. 
\begin{algorithm}[tb]
   \caption{Recursive  Extreme Studentised Deviate test}
\begin{algorithmic}
\STATE  {\bfseries Inputs:}  Dataset $\bfx =(x_1,x_2, \ldots, x_n)$
\newline  Significance level $\alpha$;
\newline Maximum number of anomalies to be tested $k$;
\newline The initial full sample mean $\bar{\tilde x}_0 = \bar x$   and $\tilde S_0^2 = \sum_{i=1}^{n} \left(x_i-\bar  x \right)^2 $;
\STATE
\STATE  {\bfseries Algorithm:}     \FOR{$j=1$ {\bfseries to} $k$}
    \STATE Identify $x^*$, the  maximum deviate in the dataset;
    \STATE Perform a recursive update of the sum of squares  using Equation 3;
\STATE Calculate the critical value $\gamma_{j-1}$ and test statistic $R_j$ using Equations $1$ and $2$;

   \IF{$R_j > \gamma_{j-1}$,}
   \STATE   Flag $x^*$ as an anomaly and attach to an anomaly vector $\bfx_A$ ;
    \STATE Recursively update $\bar{\tilde x}_{j} $, the mean of the reduced dataset;
                    \STATE Reduce the dataset by removing $x^*$;
   \ENDIF
   \ENDFOR
   \STATE
   \STATE \textbf{Outputs:} 
Anomaly vector $\bfx_A$.

\end{algorithmic}
\end{algorithm}
It enables anomaly detection in streaming data by using a rolling window approach and recursively updating the test statistics recursively as each data point arrives.
Let $x_w$ be a newly streamed data point and let $x_0$ be the datapoint that is being removed as the window rolls forward by one at time $t$. Then the sum of squares and the sample mean can be calculated at time $t+1$ by the following recursive formulae;
\begin{equation}\label{eqn:ssrecstream}
  \begin{array}{l}
 S_{t+1}^2 =  S_t^2  +\left(x_w-x_0\right)\left(x_w +x_0-2\bar x_t - \frac{x_w-x_0}{w}\right); 
\\ \bar {x}_{t+1}=\bar{ x}_t+\frac{(x_w-x_0)}{w}.
\end{array}
\end{equation}

The Recursive ESD (R-ESD) streaming algorithm for anomaly detection is presented in Algorithm 2.
It is a two stage approach. 
In the initial phase, a window of data $\bfx'$, of size $w'$, is decomposed into its seasonal $(\bfS)$, trend $(\bfT)$ and residual $(\bfepsilon)$ components, such that 
\begin{equation*}
x_t' = S_t + T_t + \epsilon_t,
\end{equation*}
where $\epsilon_t  \sim N(0,\sigma^2)$, that is, the residuals at each time step are assumed normally distributed with zero mean and variance $\sigma^2$  to be estimated. We also assume that the errors in this general model are uncorrelated in time.
These assumptions render the generalised ESD test appropriate to detect anomalies in the residuals $\bfepsilon$. Note that  the initial window size $w' \geq w$ to allow the  fit of a useful statistical model. 
We assume that in this training period, no anomalies are detected. In practice for example, an engineer would  monitor a manufacturing process carefully during this initial phase.

\begin{algorithm}[tb]
\caption{Recursive ESD Streaming Algorithm  (R-ESD)}
   \label{alg:example}
\begin{algorithmic}
   \STATE {\bfseries Inputs:} Time series data $\bfx =(x_1,x_2, \ldots, x_t)$ observed at time $t$, with streaming new observations $(x_{t+1}, \ldots)$;
\newline Initial  window size $w'$;
\newline Streaming window size $w$;
\newline Maximum number of anomalies $k$ in any given window; 
\STATE
\STATE{\bf Initial Phase:}
\STATE Define the initial training window of data by \newline
$\bfx'=(x_{t-w'+1},x_{t-w'+2}, \ldots, x_t)$;
\STATE Perform trend and seasonal decomposition of $\bfx'$ e.g. by using methods described in Section \ref{sec:grubbs};
\STATE Create forecasts e.g. using the \texttt{forecast} function \cite{forecast2}  $\bfx^f=(x^f_{t+1},\ldots)$ as far as is required for the application and/or is computationally feasible;
\STATE Define the current window of data to search for anomalies by 
$\bfx=(x_{t-w+1},x_{t-w+2}, \ldots, x_t)$ and denote the associated  stationary residuals  $\bfepsilon = (\epsilon_{1},\epsilon_{2}\dots,\epsilon_{w})$ found in the model decomposition in line 2;
\STATE Compute the initial sum of squares of the residuals $S_t^2$ using Equation \ref{eqn:ss};
 \STATE Compute the initial mean of the residuals  $\bar\epsilon_t=\sum_{i=1}^{w} \epsilon_i/w$;
 \STATE
\STATE{\bf Streaming Phase:}
   \FOR{$s=(t+1)$ {\bfseries to} $\ldots$}

\STATE Let $\epsilon_0=\epsilon_1$; 
\STATE Slide the current window of data by one observation such that $\bfx=(x_{s-w+1},x_{s-w+2}, \ldots, x_s)$;\STATE  Calculate $\epsilon_w = x_s - x^f_s$ using the forecasts found in line 2;
\STATE
Update  $S_{s}^2$ and $\bar {\epsilon}_{s}$ recursively (Equation \ref{eqn:ssrecstream});
\STATE Perform the Recursive Extreme Studentised Deviate test to detect anomalies $\bfx_{A,s}$ (Algorithm 1); 
\ENDFOR
\STATE
   \STATE {\bfseries Outputs:}  
$\{ \bfx_{A,s},t_s\}_{s=(t+1)}^{\ldots } $, where $\bfx_A,s$ is a vector of anomalies found using the R-ESD test in the $s^{th}$ window and where  $t_s$ denotes the time stamp of the last observation in the relevant window.
\end{algorithmic}
\end{algorithm}

Time series decomposition is a well-studied topic and commonly used methods are described in Chapter 6 of \cite{hyndmanbook}. 
A suitable period $p$ can be found  in the initial phase for example, by using a Fourier transformation via \texttt{periodogram}  from the \texttt{TSA} software package \cite{tsapackage}.
The \texttt{stlm} function of the \texttt{forecast} package \cite{forecast2} is then used to model the trend using LOESS and to forecast the typical behaviour of future observations as far as is required. For example, in the context of smart manufacturing, this would be the length of the production process that is about to be performed. If this is not computationally feasible or if the typical behaviour of the process is expected to change across time, a suitable model can be re-fitted as often as is deemed necessary.
The resulting model is used to calculate the residuals in the streaming window $\bfx$  and initialise the statistics required for the generalised ESD test namely, the mean of the residuals;  $\bar {\epsilon}_t=\sum_{i=1}^{w} \epsilon_i/w$ and the sum of squares of the residuals; 
\begin{equation}\label{eqn:ss}
S_t^2= \sum_{i=1}^w(\epsilon_i-\bar{ \epsilon}_t)^2.
\end{equation}

In the streaming phase,  the recursive generalised ESD test outlined in Algorithm 1 is applied to the residuals in each window. As the window slides forward by one datapoint at each iteration, fast recursive updates of $S_t^2$ and $\bar {\epsilon_t}$  at time $t$ are employed using Equation \ref{eqn:ssrecstream}  during streaming R-ESD (line 11, Algorithm 2).

\section{Results}\label{sec:results}

The R-ESD  approach is first demonstrated and compared with SH-ESD using the ``raw\_data" example presented in the \texttt{AnomalyDetection} package published on the GitHub repository \cite{twitterpackage}. The anomalies are unlabelled but it is chosen for ease of comparison with SH-ESD. A single window of data is used to make as direct as possible a comparison and the full dataset is used while streaming. All other examples are for streaming data.
The labelled machine temperature manufacturing problem introduced in Figure \ref{fig:intromachines},  Section \ref{sec:intro}  is useful to aid visualisation and understanding for a novel reader. 
Finally theYahoo EGADS benchmark data A3  (also known as TS-2) \cite{webscopedata} is used to aid comparisons with EGADS and DeepAD. 
If no prior knowledge is available selecting $wprime=10\%$ and $w=2\%$ of the data is suggested with sensitivity analysis performed around these choices.

 Figure $2$ displays the R-ESD and SH-ESD results for a single window of $4$ days of the first example. The significance level used for the generalised ESD test was $\alpha=0.05$, the default in the \texttt{AnomalyDetection} package.
The number of anomalies tested by R-ESD in the single window is $k=288$.  This is  to coincide with max\_anoms $=0.02$ as is given in the \texttt{AnomalyDetection} package, the ``maximum number of anomalies that S-H-ESD will detect as a percentage of the data" (since $k=w\times$ max\_anoms $=288$). R-ESD and SH-ESD  agree  in detecting $106$ anomalies with a further $24$  distinct anomalies detected by R-ESD and $8$ by SH-ESD. The R-ESD anomalies appear to be more convincing although there is no ground truth for this example. CPU time for SH-ESD was $0.17$ seconds, while R-ESD required $0.46$ seconds. However, while the R-ESD is computationally less efficient than SH-ESD for this example, it is statistically principled and the algorithm is designed using recursive updates of the test statistic to allow real-time anomaly detection while streaming new datapoints using  a sliding window. 

\begin{figure}[htbp]\label{fig:twitteronewindow}
\begin{center}
\centerline{\includegraphics{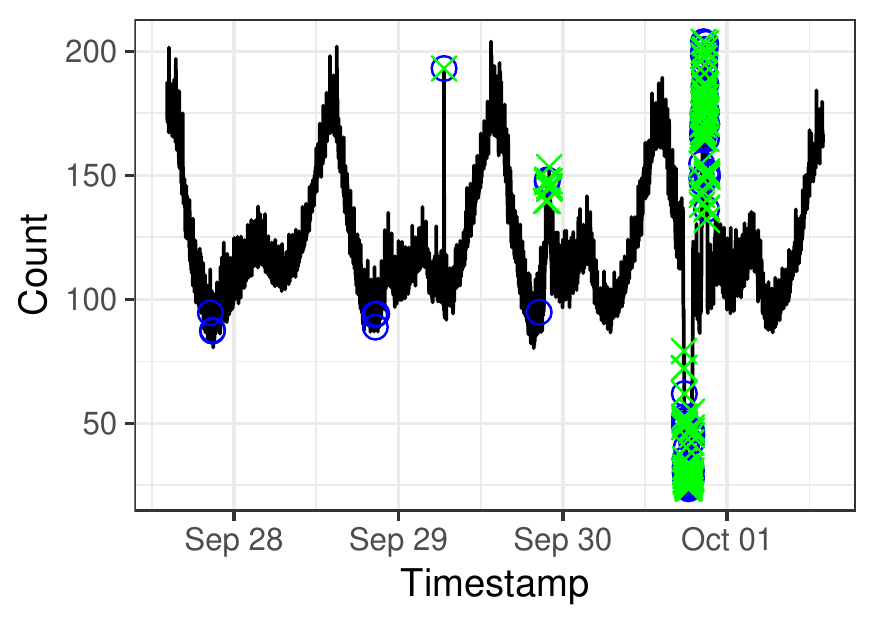}}
\caption{A single window of Twitter data: Resulting anomalies detected by R-ESD and SH-ESD are shown in green and blue respectively, with $k=288$ (and equivalently max\_anoms$=0.02$) anomalies allowed in the window. A 4 day window is used from the example given in Twitter \texttt{AnomalyDetection} package on their GitHub repository with unlabelled anomalies.}
\label{fig}
\end{center}
\end{figure}

Results for streaming windows of size $w=1440$, that is, one day of minutely data, are given in Figure $3$ and are compared to the non-overlapping window approach of SH-ESD described in \cite{twitter14} using the \texttt{longterm\_period$=$TRUE} option in the \texttt{AnomalyDetection} package. As described in Section \ref{sec:AD}, the trend in each non-overlapping window is treated by SH-ESD as a flat line corresponding to the median of the values of the data in the window. Therefore, one might expect that R-ESD is less sensitive than SH-ESD  to the choice of window size and certainly to the starting point of the algorithm. The CPU time for R-ESD to stream $7197$ data points  in this example was $65$ seconds, that is only $0.02$ seconds per window, where the window size was $w=1440$. Thus streaming is highly feasible for many applications. Moreover, the R-ESD streaming approach seems to choose more sensible anomalies than the non-overlapping window approach of SH-ESD.  Agreement occurs for $39$ anomalies with a further  $118$ and $103$ anomalies detected by R-ESD  and SH-ESD respectively.

\begin{figure}[htbp]\label{fig:twitterstream}
\begin{center}
\centerline{\includegraphics{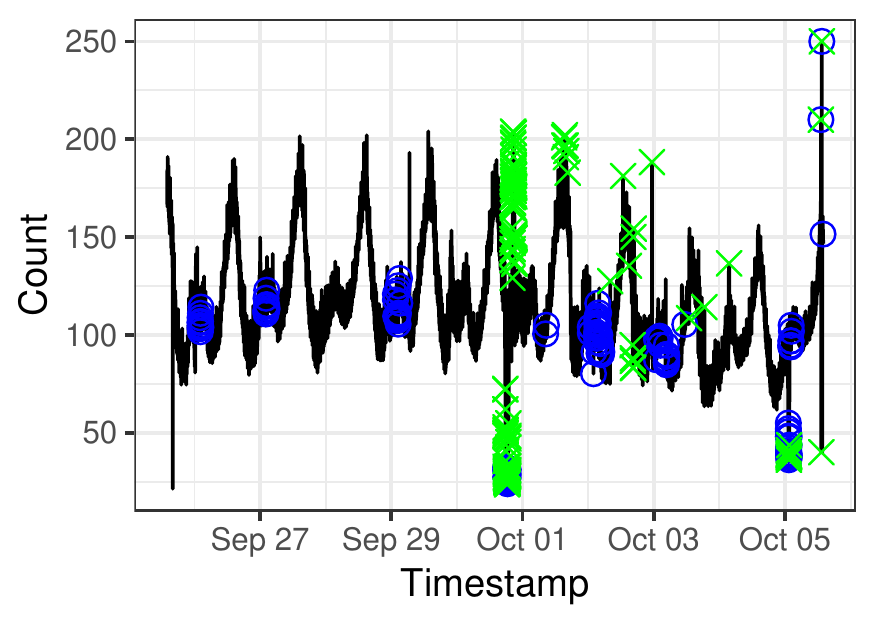}}
\caption{Streaming Twitter data: Resulting anomalies detected by R-ESD and SH-ESD are shown in green and blue respectively. The number of anomalies tested per window is $k=288$ (and equivalently max\_anoms$=0.02$ for SH-ESD) anomalies allowed in the window.}
\label{fig}
\end{center}
\end{figure}

The improved performance of R-ESD over SH-ESD is further demonstrated in the machine temperature example 
 displayed in Figure $4$ for the data described in Section \ref{sec:intro}. Here the number of anomalies per window $k=10$ was deemed appropriate in the context of manufacturing such that parts are produced within the desired specification. There are $4$ known anomalies in this dataset and these are noted as being difficult to detect, the third in particular \cite{lavin15}. 
SH-ESD failed to detect any of the $4$ known anomalies when using the non-overlapping window approach to the anomaly detection. R-ESD performs better, providing anomaly detection in advance of the first labelled anomaly and thus allowing time to alert the engineer to an anomaly in advance of the problem. Furthermore it correctly detects one of the other anomalies. In practice, the former is more useful as the engineer can intervene in advance of a machine failure.  Precision and recall are $0.004$ and $0.25$ for R-ESD. Both measures are $0$ for SH-ESD. 
 This first anomaly is not explained or discussed in \cite{lavin15}. In fact their analysis does not utilise the first portion of the dataset although it is given and labelled on the \texttt{Numenta} GitHub repository. 
The third anomaly is notoriously difficult to detect as the lead up to this anomaly is a very gradual decline in machine temperature.
In terms of CPU time, R-ESD requires approximately $10$ seconds to stream $20,000$ windows, that is only $0.0005$ seconds per window. This is slower than SH-ESD which takes $0.31$ seconds. However this is for non-overlapping windows of size $961$  rendering the two reasonably computationally similar (since $(20000/961)\times0.31\approx 6.5$ seconds). 

\begin{figure}[htbp]\label{fig:machinestream}
\begin{center}
\centerline{\includegraphics{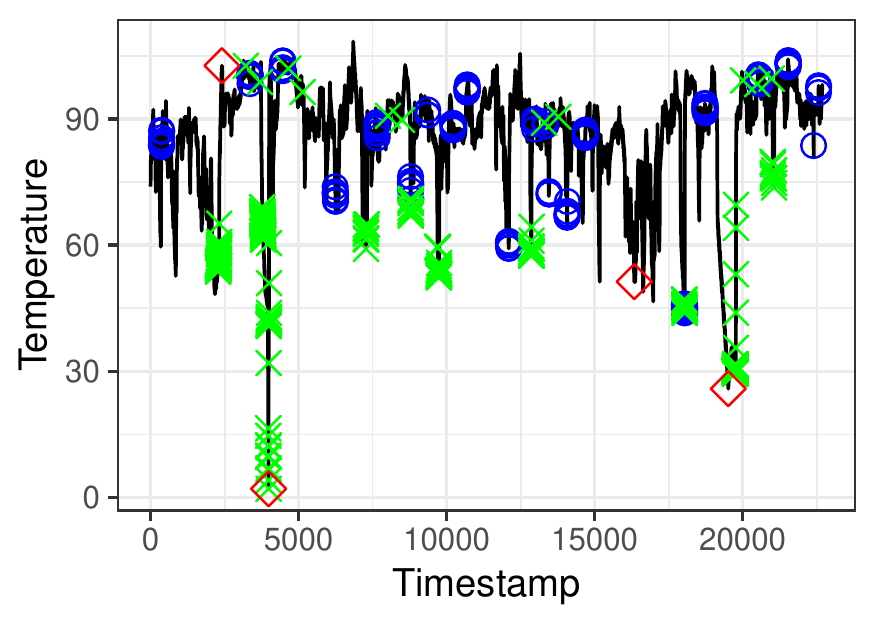}}
\caption{Streaming  machine temperature example: Anomalies detected by R-ESD and SH-ESD are shown in green and blue respectively. The number of anomalies tested per window is $k=10$ for R-ESD (and equivalently max\_anoms$=0.0004$ for SH-ESD). Known anomalies are marked in red.}
\label{fig}
\end{center}
\end{figure}

Finally, the R-ESD algorithm was implemented on the Yahoo EGADS benchmark A3 data  (also known as TS-2) \cite{webscopedata}  with results displayed in Figure $5$.  This dataset includes outliers and no change points, which is the type of dataset most suitable for good performance of both SH-ESD and R-ESD. As Latpev et al., 2015 admits, there is no best anomaly detection model for all use cases and certain algorithms are best at detecting different types of anomalies.  By comparing Figure $5$ of \cite{laptev2015egads} and the R-ESD results displayed  Figure $5$ of this article, it is clear that  all methods perform badly in comparison to R-ESD, with F1-scores of less than $0.7$ compared with greater than $0.75$ for R-ESD. 

R-ESD can also be compared with DeepADVote by examining Figure 5 below and Figure 3 of \cite{buda2018deepad}. R-ESD outperforms DeepADVote as demonstrated by a comparable but much less variable F1-score (and recall) for dataset A3, with 25th percentiles as low as ~0.3(0.2) for DeepADVote but ~0.65(0.5) for R-ESD.  R-ESD is by its nature extendable to fit any model or collection of models at every $n$th streamed datapoint with a trade-off between computational time and streaming capacity.


\begin{figure}[htbp]\label{fig:machinestream}
\centerline{\includegraphics[width=7cm,height=5cm]{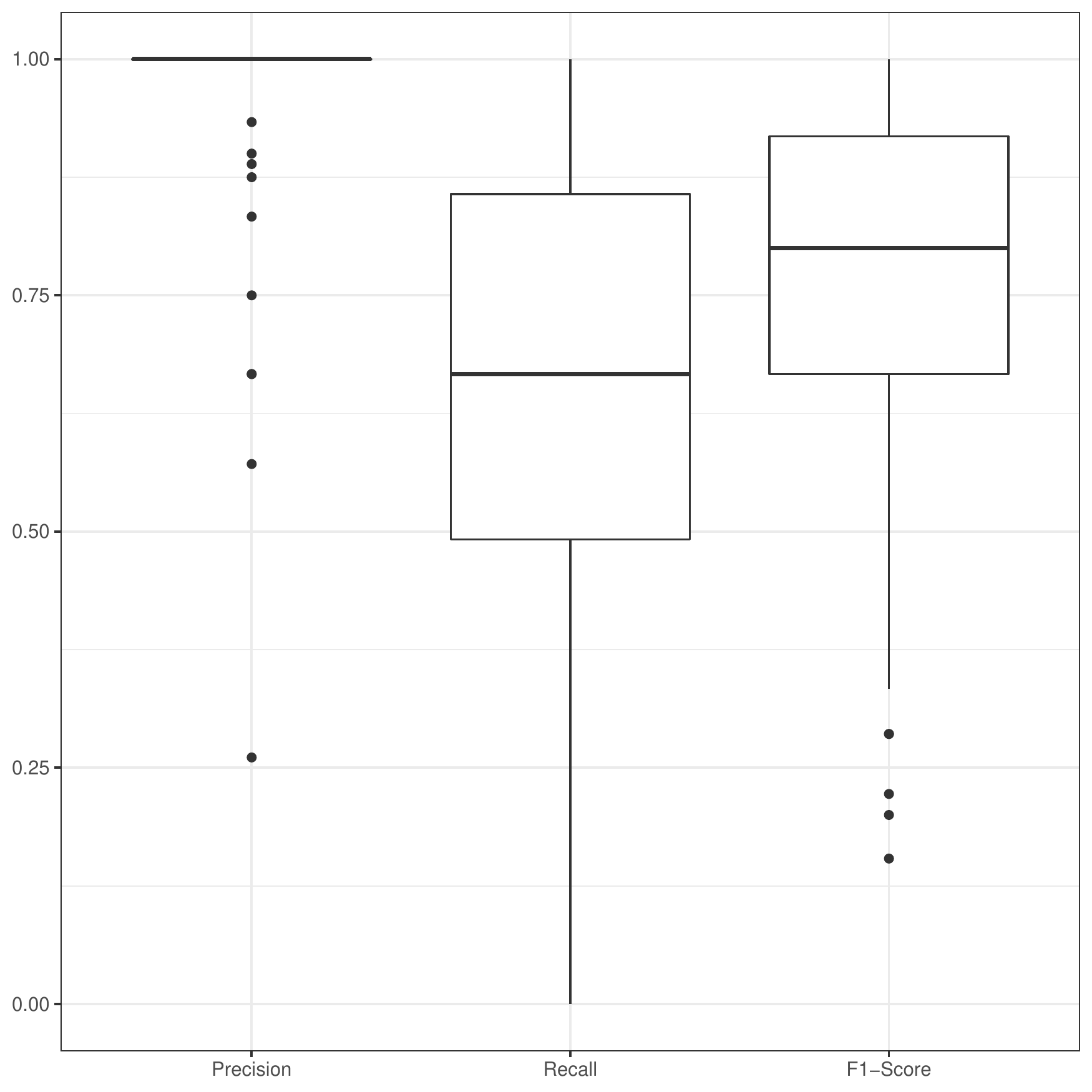}}
\caption{Boxplots of precision, recall and F1-score for the A3 Yahoo EGADS benchmark dataset.}
\end{figure}

\section{Conclusion}\label{sec:discussion}
This paper presents a novel approach to anomaly detection for streaming time series data, which typically exhibits trend and or seasonality.  The primary novelty of the R-ESD algorithm is 
 an algebraic manipulation that enables a real-time, low memory, statistically principled version of a widely used approach. 
It enables the use of multiple recursive updates within the ESD test across rolling windows of data. The major advantage is that this renders the approach feasible for streaming data. In the examples presented, computation times were as little as $0.02$ and $0.0005$ seconds per window i.e. per streamed datapoint.  
If required, computation times could be reduced further by implementing  a priority queue \cite{knuth97} to reduce memory requirements.

An extension to this algorithm would be to combine more modelling and forecasting approaches for example those give in both DeepAnT \cite{munir2018DeepAnT} and DeepAD \cite{buda2018deepad}, where one or multiple models are fit at every newly streamed datapoint. However, there would  be a considerable trade-off between computational time and forecasting accuracy which may render it unsuitable for many streaming applications. One option could be to refit the model every $n^{th}$ datapoint or every time an anomaly is detected. 
Further studies are required to extend the comparison of R-ESD using  the  Numenta benchmark tests, which explicitly reward early detection. We suspect that R-ESD will perform well by this measure given the results presented for the machine temperature example.

In summary R-ESD is a fast,   statistically principled and novel recursive approach to anomaly detection in time series data. It is highly feasible for streaming in real-time. It correctly studentises observations according to the theoretical distribution of the ESD test statistic and it outperforms the \texttt{AnomalyDetection} package, thus improving on Twitter's approach.
Beyond SH-ESD, R-ESD also outperforms EGADS \cite{laptev2015egads}, which uses a combination of many models and methods to detect as many different types of anomalies as possible and DeepAdVote, the most comparable algorithm to ours of the deep-learning based approaches to anomaly detection.
While this article focuses on the anomaly detection testing routine, 
\bibliography{resd}

\end{document}